# Stride variability measures derived from wrist- and hip-worn accelerometers

Jacek K. Urbanek[1], Jaroslaw Harezlak[2], Nancy W. Glynn[3],
Tamara Harris[4], Ciprian Crainiceanu[1], Vadim Zipunnikov[1]

1) Department of Biostatistics, Johns Hopkins Bloomberg School of Public Health, 615 N. Wolfe Street, Baltimore, MD 21205
2) Center for Aging and Population Health, Department of Epidemiology, Graduate School of Public Health, University of Pittsburgh
3) Department of Biostatistics, Indiana University School of Medicine
4) Laboratory of Epidemiology, Demography, and Biometry, National Institute on Aging

**Abstract:** Many epidemiological and clinical studies use accelerometry to objectively measure physical activity using the activity counts, vector magnitude, or number of steps. These measures use just a fraction of the information in the raw accelerometry data as they are typically summarized at the minute level. To address this problem we define and estimate two gait measures of temporal stride-to-stride variability based on raw accelerometry data: Amplitude Deviation (AD) and Phase Deviation (PD). We explore the sensitivity of our approach to on-body placement of the accelerometer by comparing hip, left and right wrist placements. We illustrate the approach by estimating AD and PD in 46 elderly participants in the Developmental Epidemiologic Cohort Study (DECOS) who worn accelerometers during a 400 meter walk test. We also show that AD and PD have a statistically significant association with the gait speed and sit-to-stand test performance.

## 1 Introduction

Accelerometers are now ubiquitous in health studies, where they are used to provide objective and reproducible proxy measurements of physical activity. Examples of such studies include both large epidemiological cohorts, such as the National Health and Nutrition Examination Survey (NHANES) [1] and the Baltimore Longitudinal Study of Aging (BLSA) [2], and clinical studies of chronic disease, such as Alzheimer's Disease [3], Multiple Sclerosis [4] and Heart Disease [5]. The primary activity measurements in these studies are usually limited to crude summaries of the 24-hour activity cycle such as the total daily activity count, vector magnitude, or number of steps. When walking is of primary scientific interest, steps-based summaries provide useful information about "how much" and "when" the person is walking, but *do not provide any information about "how" the person is walking or*

*"whether" their walking changes during the course of the day.* This type of information can be crucial in clinical and observational studies as it provides information about the intrinsic walking parameters and their associated variability. Understanding the association between these parameters and levels of fatigue and fatigability in healthy and frail populations is a major step towards identifying parameters that are intuitive, can easily be extracted from accelerometry data, and are relevant to health studies. Quantifying gait parameters and ambulatory monitoring of changes in these parameters has become increasingly important for epidemiological, clinical and rehabilitation studies.

Several approaches extracting time-dependent gait parameters were developed and successfully applied to data collected from body-worn accelerometers [6] [7] [8]. The proposed approaches demonstrated significant discriminative power in studies of clinical pathology [9] [10], fatigability [11] [12], and aging [13] [14] [15]. Stride-to-stride variability is an important gait parameter that quantifies participants' ability to maintain walking consistency and is strongly associated with motor ability [16] [17] [9]. Stride-to-stride variability has been linked to Mild Cognitive Impairment [18], dementia [19], and stroke [20]. One of the limitations of current approaches is that they are based on data obtained from accelerometers placed around the middle of the body (hip, lower back) [21] [22]. However, research has recently shifted more towards wrist-worn accelerometers such as the Actigraph Link, GENEActiv Watch, Fitbit Flex and Jawbone Up. This shift is likely due to their ease of use, increased compliance of study participants, and improvements in size and battery life. [1]. This shift raises new challenges to estimating gait parameters, as hands are involved in a much wider spectrum of activities, which results in higher complexity and increased within- and between-subject variability [23].

We propose a method to extract two measures of stride-to-stride variability: Amplitude Deviation (AD) and Phase Deviation (PD). These measures are based on the amplitude of acceleration and duration of consecutive strides, respectively. We compare the performance of AD and PD calculated based on raw accelerometry data obtained from three body locations: the hip, the left wrist, and the right wrist. We evaluate the sensitivity of AD and PD as a function of on-body placement in 46 participants of the Developmental Epidemiologic Cohort Study [24]. To benchmark AD and PD against standard accelerometry summaries and physical function tests, we evaluate their association with four measures: cadence (C), vector magnitude counts (VMC), time on Five-Times-Sit-To-Stand ($Chr_{5s}$) test [25] and usual gait speed measured on a 6 meter distance test ($Pace_{6m}$) [26].

## 2 Methods

### 2.1.1 Participants

Data were collected as a part of the Developmental Epidemiologic Cohort Study (DECOS) [27], a study of older adults in good health. Forty six participants (25 males, 21 females; age: 78±4 y.o.; BMI: 26.75±3) were selected for the analysis.

### 2.1.2 Measurement protocol

Fast paced four hundred meter walk is a standardized test measuring physical function often employed by epidemiological studies [28]. The test consisted of 20 consecutive laps, 20 meters each. For each participant we chose data collected during the second lap of the 400 meters trial as it is expected that during this lap gait parameters are more likely to represent normal gait characteristics for each participant [29]. The reasons are that during the first lap individuals may try to outperform, experience larger variability in the first part of the experiment, while effects of fatigue are less likely to occur only after 20 meters. During the task participants wore three ActiGraph GT3X+ devices (each with three orthogonal axes, sampling frequency: 80 observations/second) located on the hip, the left and right wrists. Data were collected in parallel from all three sensors and synchronized at the sub-second level between devices.

## 2.2 Data analysis

The three-axial acceleration signal was first reduced to the vector magnitude

$$r_i = \sqrt{x_i^2 + y_i^2 + z_i^2},$$

which is less sensitive to device rotation and small changes in position. Here *x, y, z* are the acceleration signals measured along the three orthogonal axes and *i* represents time. For presentation clarity we have dropped indices corresponding to participant and sensor locations.

We used fast Fourier transformation of the vector magnitude data from the second lap of the 400-meter walk trial. For each walk-trial and sensor location the mean cadence C was estimated by identifying the spectral peaks corresponding to stride-to-stride frequency. Cadence is expressed in steps-per-second and is defined as $C = f / 2$, where *f* is the frequency location of the spectral peak corresponding to stride-to-stride frequency (express in Hz). To express cadence in steps per minute, C could be multiplied by 60. An example of Fourier spectrum with estimated location of the peak of interest is presented in the top panel of Figure 1.

To estimate the duration of consecutive strides, we first extracted the walking-specific signal using a band-pass filter ranging from 0.75C to 1.25C (where C denotes cadence). The resulting signal can be interpreted as a periodic wave of slowly varying instantaneous frequency that corresponds to durations of consecutive steps. Duration of each stride length was estimated by localizing even zero-crossing points (see fig.2 – middle panel).

## 2.3 Strides synchronization

After the data transformation steps described in Section 2.2, stride-specific patterns were time-synchronized. To do this, data was first interpolated using splines and then linearly aligned to 0 to 1, where location 0 marks the beginning of the stride cycle while 1 marks its end (see fig.1 – bottom panel). This transformation allows the estimation of the average stride profile for each participant and sensor location. The average value of the amplitude of the acceleration signal within the stride-cycle is expressed as:

$$A_\phi = \frac{1}{M} \sum_{j=1}^{M} r_{\phi j},$$

where $\phi$ is an index of a stride phase ranging from 0 to 1 and $j$ denotes the stride index. Examples of average stride acceleration profiles for all three locations are displayed in figure 2.

### 2.3.1 Stride variability measures

We focus on two measures of stride variability that reflect differences in amplitude and phase. We define the stride-to-stride amplitude deviation (AD) as the mean standard deviation of synchronized activity count profiles:

$$AD = \frac{1}{N} \sum_{i=1}^{N} \sqrt{\frac{1}{M} \sum_{j=1}^{M} (r_{ij} - A_i)^2},$$

where N is the number of samples for each synchronized profile, M is the total number of consecutive profiles, and $A_i$ denotes the average stride profile at sample time $i$. We also define the stride-to-stride phase deviation (PD) as the standard deviation of estimated durations of strides:

$$PD = \sqrt{\frac{1}{M} \sum_{j=1}^{M} \left(S_j - \frac{1}{f}\right)^2},$$

where $S_j$ denotes the duration of $j$-th stride expressed in seconds and $1/f$ is the average stride length estimated in Section 2.2.

### 2.3.2 Additional measures

Several additional parameters were estimated. In particular, we estimated the mean cadence, C, for each participant and device location. Cadence was calculated as the inverse of the average stride length multiplied by a factor of two and expressed in steps-per-minute. We have also computed the average vector magnitude count

(VMC) for each sensor location. VMC is defined as the mean absolute deviation of the acceleration signal:

$$VMC = \frac{1}{T}\sum_{t=1}^{T}\left|r_t - \frac{1}{T}\sum_{s=1}^{T}r_s\right|,$$

where T denotes the total number of samples for each gait acceleration signal.

### 2.4 Dependent variables

We compare the estimated measures, AD, PD, C and VMC, with performance measures obtained from standardized tests of physical function administered in the DECOS study. We focused on the five-times-sit-to-stand ($Chr_{5s}$) test (mean = 0.4, SD = 0.1) [25] and the usual gait speed measured by a 6-meter walk ($Pace_{6m}$) test (mean = 1.1, SD = 0.2) [26].

### 2.5 Statistical analysis

The association between each stride characteristic and dependent variable was evaluated using linear regression models. Each model was adjusted for age. The estimated parameters and p-values for the null hypothesis of no association are presented in table 2. We have also calculated the correlation between measurements using the Spearman's rank correlation ρ. Figure 3 displays XY-plots of the proposed measures and the estimated correlation coefficients.

## 3 Results

We compared the sensitivity to body location of the four accelerometry-derived gait parameters then studied their association with standard measures of physical function.

### 3.1 Hip vs. Wrist Placement

Figure 3 provides the pair-wise scatterplots (the upper triangle) and the corresponding Spearman's rank correlations (the lower triangle) for each of the four gait parameters. Stride-to-stride AD exhibits a relatively high left vs. right wrist correlation (ρ = 0.67), and a lower left-wrist versus hip (ρ = 0.52) and right-wrist versus hip correlations (ρ = 0.36). Stride-to-stride PD indicates higher left versus right wrist correlation (ρ = 0.67), and lower left-wrist versus hip (ρ = 0.36) and right-wrist versus hip correlations (ρ = 0.41). The estimated PD values using wrist accelerometry tend to be larger than PD values obtained using hip accelerometry.

VMC exhibited high correlation between the two wrists (ρ = 0.78) and between the hip and either wrist (ρ = 0.79 and 0.73, respectively). The VMC values estimated from the hip-worn sensor were slightly higher than those obtained from wrist-worn sensors. Cadence indicated much higher correlation across all three locations ($\rho$ = 0.98, 0.99, 0.99) without significant bias.

## 3.2 Association with physical function tests

Table 2 provides the regression coefficients and the p-values of the tests for the null hypothesis of no-association between the newly proposed measures of gait and standard measures of physical function. Models are adjusted for age and were applied separately for each sensor placement: hip (top panel), left wrist (middle panel), and right wrist (bottom panel). AD was significantly associated with $Chr_{5s}$ for all three locations (p < 0.01) and was significantly associated with $Pace_{6m}$ for the hip-located sensor (p = 0.014). No significant relationship was identified between PD and any of the dependent variables. Walking-related VMC was strongly associated with both $Chr_{5s}$ and $Pace_{6m}$ for all three locations (p < 0.05). Cadence estimated from data collected at the hip was significantly associated with both $Chr_{5s}$ (p = 0.043) and $Pace_{6m}$ (p = 0.037). Cadence estimated from data collected at the right wrist was not statistically associated with $Pace_{6m}$ (p = 0.055) and $Chr_{5s}$ (p = 0.097). We have found statistically significant associations between the cadence estimated from data collected at the left wrist and $Pace_{6m}$ (p = 0.032) but not with $Chr_{5s}$ (p = 0.058).

## 4  Discussion

We have proposed and examined two measures of stride-to-stride variability, Amplitude Deviation and Phase Deviation, estimated from raw accelerometry data collected during walking. AD was significantly associated with the standardized performance tests, $Chr_{5s}$ and $Pace_{6m}$ across all three locations. PD was not associated with these tests. As our population represents healthy aging participants with no clinical diagnosis, PD has the potential for use in specific clinical populations.

When we compared AD and PD across the three on-body locations, gait parameters were more correlated for data collected from the left and the right wrists and less correlated for data collected from the wrist versus hip. The cadence (C) and the vector magnitude count (VMC) are consistently associated with $Chr_{5s}$ and $Pace_{6m}$.

A limitation of all wrist-based accelerometry-derived gait parameters is that sensors record arm movement, which are proxies of walking strides [30]. Moreover, in the free-living environment one would expect long periods of walking without arm swinging (e.g. because hands are in the pocket or handling a phone). However, our results indicate the accelerometry-derived parameters can be useful in studies of normative aging and may offer opportunities for monitoring temporal gait variability in impaired populations. We conclude that raw accelerometry data contains more detailed information about gait characteristic than the currently used actigraphy summaries.

## 5 Funding

This research was supported by NIH grant RC2AG036594, Pittsburgh Claude D. Pepper Older Americans Independence Center, Research Registry and Developmental Pilot Grant -NIH P30 AG024826 and NIH P30 AG024827 and National Institute on Aging Professional Services Contract HHSN271201100605P. This project was also supported, in part, by the Intramural Research Program of the National Institute on Aging.
This research was supported by the NIH grant RO1 NS085211 from the National Institute of Neurological Disorders and Stroke, by the NIH grant RO1 MH095836 from the National Institute of Mental Health and by NIH grant 1R01 HL123407 from National Hearth, Lung and Blood Institute.
Dr Harezlak's research was supported in part by the NIMH grant R01MH108467.

# Figures and tables

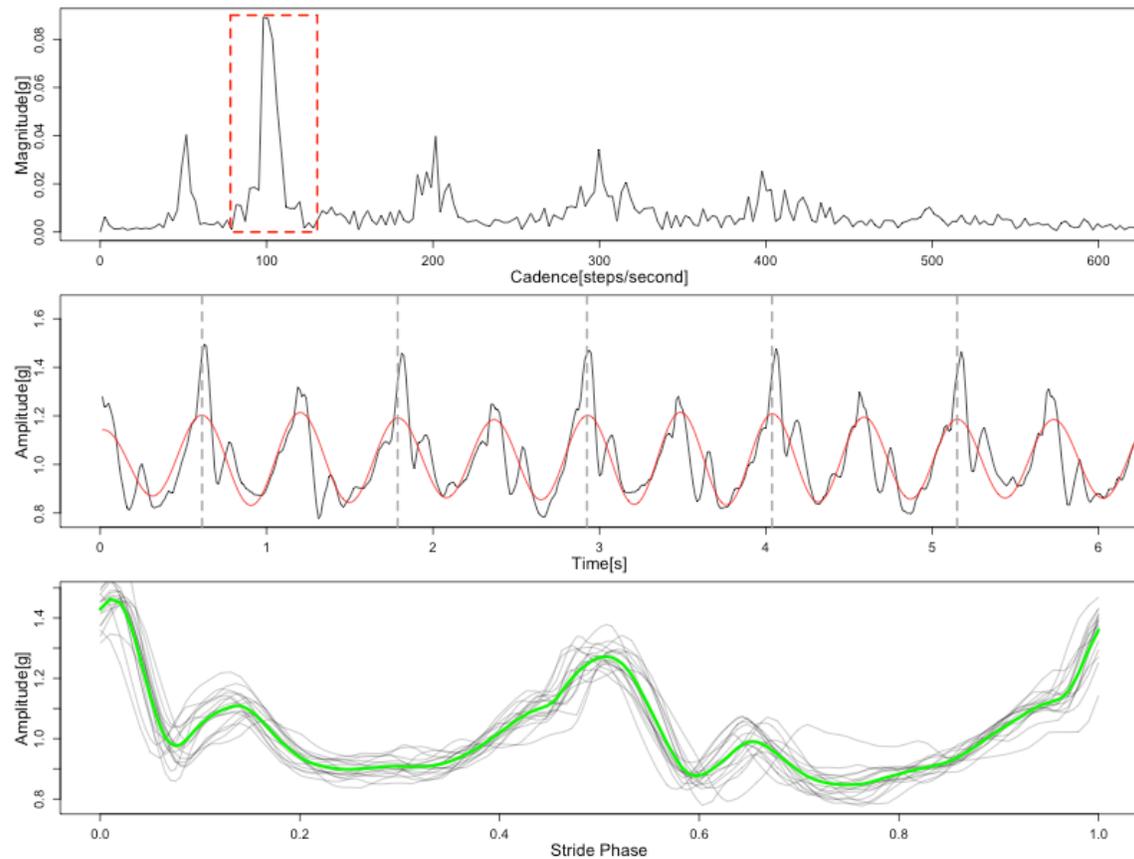

Fig.1. Stages of strides synchronization process. Top panel – Fourier spectrum of gait acceleration signal. Red box marks frequency range corresponding to step-to-step frequency. Middle panel – Time view of gait acceleration signal (black line) and signal after filtration (red line). Bottom panel - synchronized stride profiles (gray lines) and resulting average stride profile (green line).

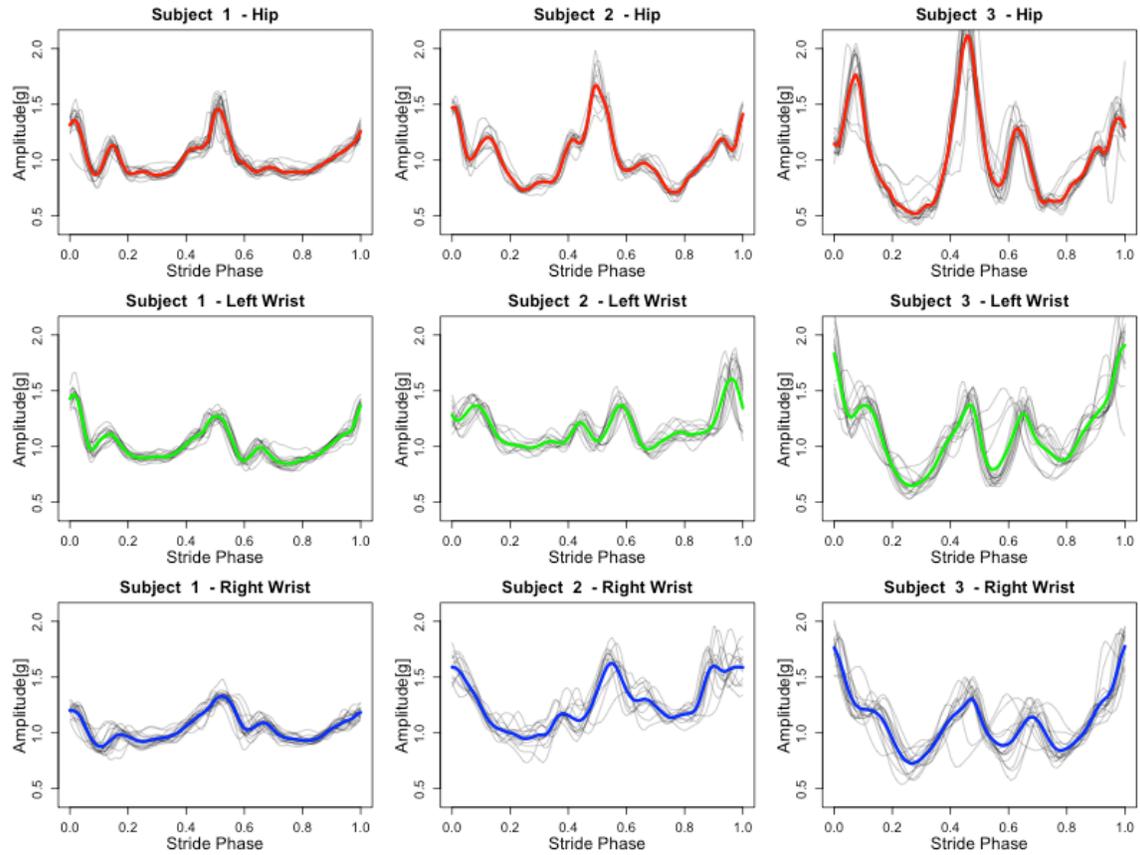

Figure.2. Exemplary stride profiles for three participants and three sensor locations. Gray lines correspond to registered stride profiles. Color lines represent averaged stride profiles. Red – hip, green – left wrist, blue – right wrist.

|  | Hip | Left Wrist | Right Wrist |
|---|---|---|---|
| VMC | 0.27 (0.09) | 0.25 (0.14) | 0.26 (0.14) |
| AD | 0.08 (0.03) | 0.11 (0.05) | 0.11 (0.05) |
| Cadence | 122.97 (10.90) | 122.93 (10.99) | 122.80 (10.92) |
| PD | 0.02 (0.009) | 0.04 (0.029) | 0.05 (0.035) |

Table 1. Means and standard deviations (in brackets) of calculated values.

| **Hip** | VMC | AD | Cadence | PD |
|---|---|---|---|---|
| Chr$_{5s}$ | 0.73(5.85e-06) | 1.54(0.000926) | 0.16(0.04348) | -1.20(0.450201) |
| Pace$_{6m}$ | -5.15(0.000369) | -10.20(0.0143) | -1.35(0.0366) | 20.00(0.1336) |

| **Left wrist** | VMC | AD | Cadence | PD |
|---|---|---|---|---|
| Chr$_{5s}$ | 0.31(0.002746) | 0.77(0.005161) | 0.13(0.09671) | 0.43(0.345273) |
| Pace$_{6m}$ | -2.01(0.0272) | -2.47(0.325) | -1.24(0.0552) | 1.12(0.7768) |

| **Right wrist** | VMC | AD | Cadence | PD |
|---|---|---|---|---|
| Chr$_{5s}$ | 0.25(0.015096) | 0.85(0.001073) | 0.15(0.05840) | 0.65(0.15522) |
| Pace$_{6m}$ | -1.95(0.0299) | -2.66(0.268) | -1.37(0.0323) | 0.57(0.8695) |

Table 2. Slope coefficients and p-values (in brackets) for linear regression fit using proposed values. Shaded fields marks values above significance levels.

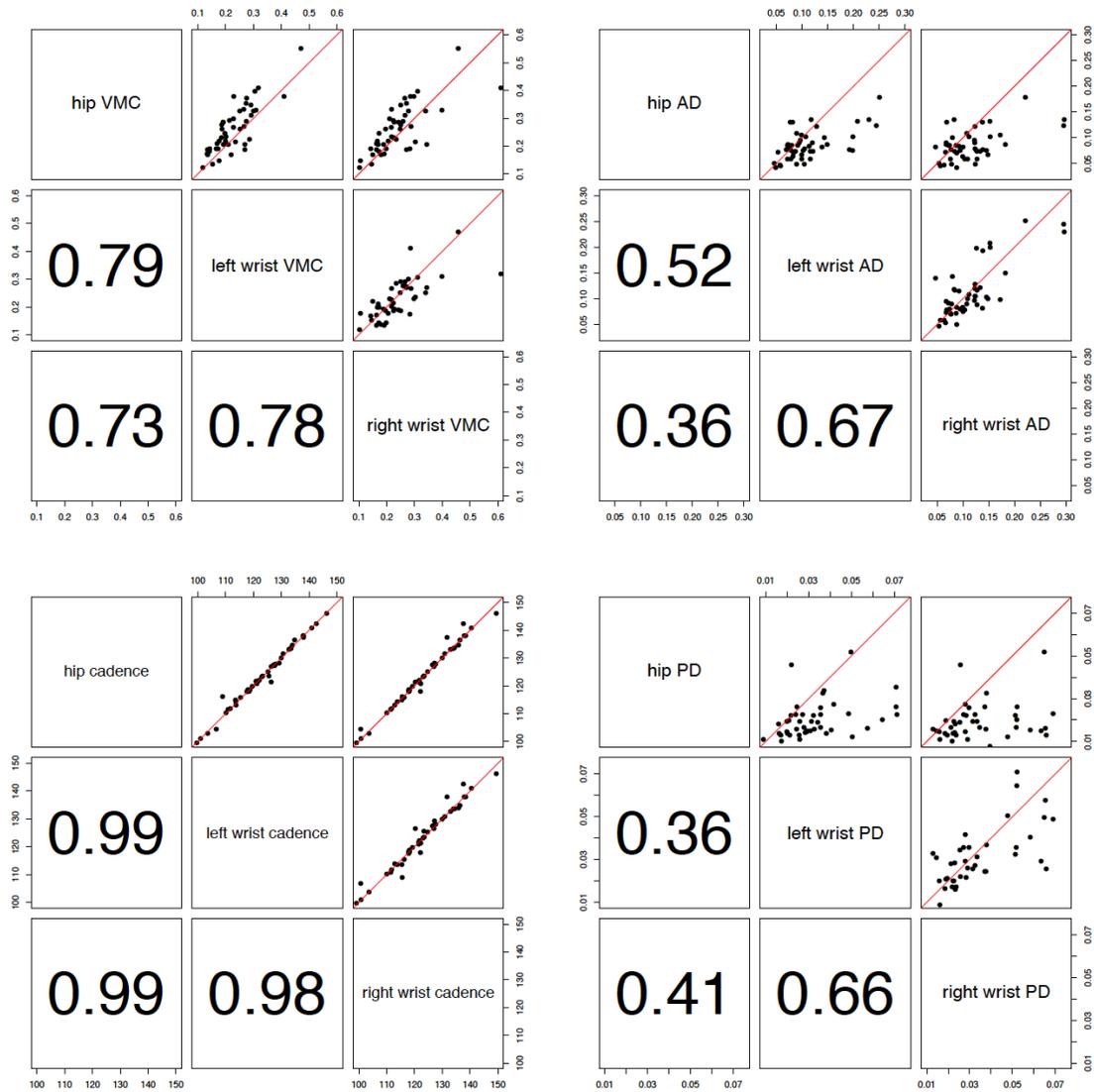

Fig.3. Pair plots displaying proposed measures against each other with red identity lines. Panels below diagonals return Spearman's rank correlation coefficients.